\documentclass[11pt]{article}

\usepackage{graphicx}
\usepackage{epstopdf}
\usepackage{amsmath}
\usepackage{amssymb}
\usepackage{amsfonts}
\usepackage{amsthm}
\usepackage[usenames]{color}
\usepackage{array}
\usepackage{mathrsfs}
\usepackage{epic,eepic,pstricks}
\usepackage{rotating}

\usepackage[
      colorlinks=true,
      linkcolor=blue,
      urlcolor=blue,
      filecolor=blue,
      citecolor=red,
      pdfstartview=FitV,
      pdftitle={},
      pdfauthor={},
      pdfsubject={},
      pdfkeywords={},
      pdfpagemode=None,
      bookmarksopen=true
]{hyperref}

\usepackage{epsfig}

\usepackage{hyperref}

\DeclareMathOperator{\Tr}{Tr}

\newcommand{\bea}{\begin{eqnarray}}
\newcommand{\eea}{\end{eqnarray}}
\newcommand{\ba}{\begin{array}}
\newcommand{\ea}{\end{array}}
\newcommand{\ee}{\end{equation}}

\numberwithin{equation}{section}

\begin{document}

\begin{flushright}
\texttt{\today}
\end{flushright}

\begin{centering}

\vspace{2cm}

\textbf{\Large{
Logarithmic Correction to BMSFT Entanglement Entropy  }}

  \vspace{0.8cm}

  {\large   Reza Fareghbal, Pedram Karimi }

  \vspace{0.5cm}

\begin{minipage}{.9\textwidth}\small
\begin{center}

{\it  Department of Physics, 
Shahid Beheshti University, 
G.C., Evin, Tehran 19839, Iran.  }\\

  \vspace{0.5cm}
{\tt  r$\_$fareghbal@sbu.ac.ir, pedramkarimie@gmail.com}
\\ $ \, $ \\

\end{center}
\end{minipage}


\begin{abstract}
 Using Rindler method we derive the logarithmic correction to the entanglement entropy of a two dimensional BMS-invariant field theory (BMSFT). In particular, we present a general formula for extraction of the logarithmic corrections to both the thermal and the entanglement entropies. We  also present a CFT formula related to the logarithmic correction of the BTZ inner horizon entropy which results in our formula after taking appropriate limit.

\end{abstract}

\end{centering}

\newpage



\section{Introduction}
 
  One of the lessons of the  AdS/CFT correspondence \cite{Maldacena:1997re} is that the asymptotic symmetry of the asymptotically AdS spacetimes in d+1 dimensions is the same as conformal symmetry in one dimension lower. One can use this idea to generalize the gauge/gravity duality beyond the AdS/CFT correspondence. Accordingly , in any non-AdS/non-CFT correspondence, the symmetry of the dual field theory should be the same as the asymptotic symmetry of the gravity solutions. 
  
  The asymptotic symmetry of asymptotically flat spacetimes were known long before Maldacena's conjecture. Such symmetries are known as BMS, which were first found at  null infinity of four-dimensional asymptotically flat spacetimes \cite{BMS}, and later were generalized to the three-dimensional case \cite{Ashtekar:1996cd}. Recently, Barnich and his collaborates \cite{Barnich:2006av} have  shown that imposing just locally well-definiteness condition is enough to enhance both the translation and the rotation symmetry of the Poincare group at null infinity to an infinite-dimensional symmetry group. In this work BMS group refers to this infinite-dimensional group which consists of super-rotation and super-translation generators.   
  
In one  dimension lower than gravity theory, the BMS algebra is given by an ultra-relativistic contraction of the conformal algebra. Thus BMS symmetry can be  the symmetry of a  lower dimensional field theory. It is proposed that the holographic dual of d+1-dimensional asymptotically flat spacetimes are ultra-relativistic d-dimensional BMS-invariant field theories (BMSFTs) \cite{Bagchi:2010zz},\cite{Bagchi:2012cy}.    
 
 In flat-space holography, BMSFT entanglement entropy and its holographic dual were first studied in \cite{Bagchi:2014iea} and the entanglement entropy of a two-dimensional BMSFT was computed by a method similar to the CFT \cite{Calabrese:2004eu}. A holographic interpretation of the BMSFT entanglement entropy in \cite{Bagchi:2014iea} (similar to the Ryu-Takayanagi proposal in the context of AdS/CFT correspondence \cite{Ryu:2006bv}) is given using Chern-Simons formulation of three-dimensional flat-space gravity \cite{Witten:1988hc}. Other aspects of BMSFT entanglement entropy has been studied in  \cite{Hosseini:2015uba},\cite{Basu:2015evh}. A remarkable progress in this subject has been achieved recently in  \cite{Jiang:2017ecm} where Rindler method \cite{Casini:2011kv} is used to derive not only the BMSFT entanglement entropy formula but also the holographic description in terms of some curves length in the bulk theory. The central idea of the Rindler method is to find local unitary transformations which map entangled states to the thermal states in the field theory. Then one can use the thermal entropy formula and find the entanglement entropy. 

Calculation of thermal entropies in BMSFTs is performed using Cardy-like formula first  introduced in \cite{Bagchi:2012xr}.  Using Rindler method, the entanglement entropy is given by  Cardy-like formula. Similar to the Cardy-formula in CFT , the saddle point approximation is used to derive this formula. Thus, it is possible to improve approximation and find possible corrections of this formula. In  \cite{Bagchi:2013qva}, employing a method first used for the Cardy formula in \cite{Carlip:2000nv}\footnote{ We note however that  the Cardy (high temperature) limit is not always reliable for extracting the black hole entropy \cite{Sen:2012cj}. On the other hand, in two dimensional CFTs the logarithmic corrections  are only universal in the Cardy  regime. From the dual  three dimensional gravity perspective, the universality of the correction is usually because heat kernels do not contain logarithmic terms \cite{Sen:2012dw}. Thus   the logarithmic correction to the BMSFT thermal and entanglement entropies are universal in the limit that BMSFT Cardy-like  formula is reliable. }, logarithmic correction to the  Cardy-like formula has been derived. 
 
In this paper, we use the logarithmic correction to Cardy-like formula along with Rindler method to find the logarithmic correction of BMSFT entanglement entropy. Similar to the leading term, this correction depends on central charges of BMS algebra, the interval of the sub-system and the cut-off. However, the interesting point is that we can rewrite the corrections in a universal form:
 \begin{equation}\label{our formula}
S=S_0-3\log\left(C_M^{\frac13}\dfrac{\partial S_0}{\partial C_L}\right)
\end{equation}
where $S_0$ is the leading term and $C_L$ and $C_M$ are the central charges of BMS algebra. $S$ can be both of the thermal entropy (which is given by the Cardy-like formula) and the entanglement entropy. This formula works for all  BMSFTs on plane or cylinder at zero or finite temperature.

 The entanglement entropy of the boundary theory can be used to   reconstruct the bulk dynamics \cite{Faulkner:2013ica, Lashkari:2013koa}. In this view, the  logarithmic correction of the entanglement entropy  might be helpful to find  quantum correction to the Einstein field equations. This paper is the first step on this road. The idea is to use the first law of the entanglement entropy as the analogue of the first law of (black hole) thermodynamics    \cite{Faulkner:2013ica, Lashkari:2013koa, Allahbakhshi:2013rda}.  The better understanding of BMSFT modular hamiltonian might allow us to achieve not only the classical bulk dynamics but also the quantum correction to the Einstein equation without the cosmological constant.

One approach to study Flat/BMSFT is to take limit from the AdS/CFT calculations. According to the proposal of \cite{Bagchi:2012cy}, the flat space limit in the gravity side corresponds to taking an ultra-relativistic limit from the CFT calculations. Hence one can find all  BMSFT formulas by taking limit from a CFT counterpart. It was shown in  \cite{Fareghbal:2014qga} and \cite{Riegler:2014bia} that the Cardy-like formula of BMSFT is given by taking limit from  a formula in the  CFT which is related to the inner horizon entropy of the gravity solution. If we assume the same relation for the logarithmic correction, the logarithmic term in \eqref{our formula} is related to a logarithmic correction in the entropy of the inner horizon. We propose a suitable logarithmic correction in the inner horizon formula which corresponds to the logarithmic term of \eqref{our formula} after taking limit. Using logarithmically corrected inner and outer horizon entropy formulas we can calculate their multiplication. The observation is that the previously known result, the multiplication being  mass independent for the Einstein gravity \cite{Larsen:1997ge} (see also \cite{Detournay:2012ug})
, is violated in the presence of the logarithmic terms.

The structure of this paper is as follows:  In section two we review the results of \cite{Jiang:2017ecm} which uses the Rindler method to derive the BMSFT entanglement entropy and its holographic description.  In section three we review the derivation of BMSFT Cardy-like formula and its logarithmic correction. In section four we put together the results of section two and three to derive the logarithmic correction of BMSFT entanglement entropy. Section five is devoted to the derivation  of the BMSFT entanglememnt entropy formula by taking flat-space limit.  
 
\section{Entanglement Entropy of BMSFT using Rindler method}\label{EE of BMSFT}

One of the advantages of the Rindler method is the convenience it provides for the calculation of entanglement entropy in the context of gauge/gravity duality. In this view, the thermal entropy of the boundary theory is mapped to the horizon entropy of the black hole (object) in the gravity side. Thus, one can use it to prove Ryu-Takayanagi (RT) formula for the holographic entanglement of CFTs. Moreover, since many thermal properties of the gravitational systems are known for the non-AdS cases, we expect that the Rindler method gives a lot of insights to find the analogue of RT formula for the dualities beyond the  AdS/CFT correspondence. 

In Rindler method, the asymptotic symmetries play an essential role.
 We are interested in  BMSFTs which are proposed to be the holographic dual of asymptotically flat spacetimes.
   In our case,  a Rindler transformation is of the form of BMS transformation and final coordinates are invariant under specific thermal identifications.
  These transformations act like unitary operators $U_{\mathcal{R}}$ on the fields and map the reduced density matrix in the subregion $\mathcal{A}$ (which we want to calculate the entanglement entropy for)  to a thermal density matrix in the interval $\mathcal{B}$,
  \begin{equation}
   \rho_{\mathcal{A}} = U_{\mathcal{R}}\, \rho_{\mathcal{B}}\, U_{\mathcal{R}}^{-1}.
  \end{equation}
  Since the unitary transformations do not change the entropy, the thermal entropy of the subregion $\mathcal{B}$ is the same as entanglement entropy of the subregion $\mathcal{A}$.

A two-dimensional BMSFT has the following symmetry \cite{Barnich:2012xq}:
\begin{align}
\nonumber \label{s1:5}
 \tilde{u} &= \partial_{\phi}{f(\phi)} u +g(\phi),
  \\  
  \tilde{\phi} &= f(\phi), 
\end{align}
where $f(\phi)$ and $g(\phi)$ are arbitrary functions of the original coordinate $(u,\phi)$. The BMS algebra is given by using  infinitesimal BMS transformation as
\begin{align}
\nonumber [L_n,L_m]&=(n-m)L_{n+m}+{C_L\over12}n(n^2-1)\delta_{n+m,0},\\
\nonumber [L_n,M_m]&=(n-m)M_{n+m}+{C_M\over12}n(n^2-1)\delta_{n+m,0},\\
  {[}M_n,M_m]&=0.
\end{align}
where for a BMSFT on a plane with coordinates $(u,\phi)$ the generators $L_n$ and $M_n$ are given by
\begin{align}
\label{s1:7}
\nonumber L_{n} &= -u (n+1) \phi^n \partial_{u} - \phi^{n+1} \partial_{\phi}
 \\
 M_{n} &= \phi^{n+1}\partial_{u}.
\end{align}
The global part of this algebra is identified with $n=0,\pm1$.

 A Rindler transformation is of the form $\tilde x=T(x)$, but it should be invariant under some imaginary identification (thermal identification) of the new coordinate $\tilde x^i\sim \tilde x^i+i\tilde \beta ^i$. Moreover, vectors $\partial_{\tilde x^i}$ annihilate the vacuum and hence should be written as the linear combination of the global part of the BMS algebra:
\begin{equation}
 \label{s1:6}
 \partial_{\tilde{x}^i} = \sum_{n=-1}^{1}{( b_{n} L_{n} +d_{n} M_{n} )}.
\end{equation}
Using \eqref{s1:5} and \eqref{s1:6} we conclude that 
\begin{equation}
\label{s1:11}
 \partial_{\phi}{f(\phi)} = \frac{1}{Y}, \qquad \partial_{\phi}{g(\phi)} = -\frac{T}{Y^2},
\end{equation}
where
\begin{align}
\label{s1:9}
 Y &= - b_{-1} -b_{0} \phi -b_{1} \phi^2,
 \\
 \label{s1:10}
 T &= d_{-1} + d_{0} \phi + d_{1} \phi^2.
\end{align}
Solutions to \eqref{s1:11} determine vector $ \partial_{\tilde{x}^i}$.
It is assumed that  entangled region is given by $\mathcal{A} = \{ (\frac{-l_{\phi}}{2} , \frac{-l_{u}}{2}) \cup (\frac{l_{\phi}}{2} , \frac{l_{u}}{2})\} $. We can use the following  constraints  to find vector $ \partial_{\tilde{x}^i}$ and the geometric (modular) flow $ k_t = -\tilde{\beta_{\phi}} \partial_{\tilde\phi} + \tilde{\beta_{u}} \partial_{\tilde u} $:
\\
\begin{enumerate}
\item  The finite interval on $\mathcal{A}$ should be  mapped  to the infinite interval on $\mathcal{B}$,
\begin{align}
 \label{s1:12}
 ( \frac{-l_{\phi}}{2} , \frac{-l_u}{2} ) &\rightarrow ( -\infty , -\infty ),
 \\
 \label{s1:13}
  ( \frac{l_{\phi}}{2} , \frac{l_u}{2} ) &\rightarrow ( \infty , \infty ).
\end{align}
\\
\item  The origin of the entangled interval is  mapped to the origin of the thermal interval
\begin{equation}
 \label{s1:14}
 (0, 0) \rightarrow (0,0)
\end{equation}
\item  The thermal interval (tilde coordinate) obeys a thermal identification of the following form
\begin{equation}
\label{s1:15}
 ( \tilde{\phi} , \tilde{u} ) \sim ( \tilde{\phi} + i \tilde{\beta}_{\phi} ,\tilde{u} - i \tilde{\beta}_{u} ) .
\end{equation}
\item  The modular flow $k_t$ vanishes on the boundary of the entangled region
\begin{align}
\label{s1:16}
 k_t(\partial{\mathcal{A}} ) = 0 \Rightarrow \begin{cases}
                                              k_t( \frac{-l_\phi}{2} ,\frac{-l_u}{2} ) = 0,
                                              \\
                                              k_t( \frac{l_\phi}{2} ,\frac{l_u}{2} ) =0 
                                             \end{cases}
\end{align}
\end{enumerate}
Using these conditions,  \cite{Jiang:2017ecm}  completely determines the Rindler transformation and  modular flow of a BMSFT on the plane as below:
\begin{align}
 \label{s1:17}
 \tilde{\phi} &= \frac{ \tilde{ \beta}_{\phi} }{ \pi } \tanh^{-1}{ \frac{2 \phi }{ l_{\phi}} } ,
 \\
 \label{s1:18}
 \tilde{u} + \frac{ \tilde{\beta}_{u} }{ \tilde{\beta}_{\phi} } \tilde{\phi} &= \frac{2 \tilde{\beta}_{\phi} ( u l_{\phi} - l_{u} \phi)}{\pi (l_{\phi}^2 - 4 \phi^2)},
\end{align}

\begin{equation}
 \label{s1:19}
 k_t = -\tilde{\beta}_{\phi} \partial_{\phi} + \tilde{\beta}_{u} \partial_{u} =\frac{- \pi}{2 l_{\phi}} \left( \left(l_{\phi}^2 - 4 \phi^2\right)\partial_{\phi} + \left(l_u l_{\phi} +4 \frac{lu}{l_{\phi}} \phi^2- 8 u \phi\right)\partial_{u} \right) 
\end{equation}

 For a BMSFT with identification of coordinates as 
\begin{equation}
\label{s1:24}
 (\tilde{u} , \tilde{\phi} ) \sim ( \tilde{u} + i \bar{a} , \tilde{\phi} - i a) \sim ( \tilde{u} + 2 \pi \bar{b} , \tilde{\phi} - 2 \pi b), 
\end{equation}
 The degeneracy of states is given by a Cardy-like formula \cite{Bagchi:2012xr}, \cite{Jiang:2017ecm}, \cite{Basu:2017aqn} (see next section).
 \begin{equation}
 \label{s1:25}
  S_{\bar{b}|b}(\bar{a}|a) = \frac{- \pi^2}{3} \left(C_{L} \frac{b}{a} +C_{M} \frac{(\bar{a} b - a \bar{b} )}{a^2}\right). 
 \end{equation}
Using \eqref{s1:25}, the entanglement entropy of a BMSFT on the plane for the interval    $\mathcal{A} = \{ (\frac{-l_{\phi}}{2} + \epsilon_{\phi} , \frac{-l_{u}}{2} + \epsilon_{u} ) \cup (\frac{l_{\phi}}{2} -\epsilon_{\phi} , \frac{-l_{u}}{2} - \epsilon_{\phi} )\} $
becomes \cite{Jiang:2017ecm}
\begin{equation}
 S_{EE} = \frac{C_{L}}{6} \log{ \frac{l_{\phi}}{\epsilon_{\phi}} } + \frac{C_{M} }{6} \left( \frac{l_{u}}{l_{\phi} }- \frac{\epsilon_{u}}{\epsilon_{\phi}}\right),
\end{equation}
where $\epsilon_u$ and $\epsilon_\phi$ are ultraviolet  cut-offs in $u$ and $\phi$ coordinates. Similarly the entanglement entropy for finite temperature BMSFT  on the  cylinder has been calculated in \cite{Jiang:2017ecm}.

 \section{Logarithmic Correction to  Cardy-like Formula of BMSFT}
The entropy of CFT thermal states is calculated using Cardy formula. Using the saddle point approximation, it is possible to find a similar formula for the degeneracy of thermal states in a BMSFT \cite{Bagchi:2012xr}, \cite{Jiang:2017ecm}, \cite{Basu:2017aqn}. Due to the Rindler method, any correction to Cardy-like formula has  influence on the entanglement entropy formula. In this section, we review the logarithmic correction to Cardy-like formula \cite{Bagchi:2013qva} and then use it to find the logarithmic correction to the entanglement entropy. Method of \cite{Bagchi:2013qva} is based on  \cite{Carlip:2000nv} that introduces the first order logarithmic correction to the Cardy formula (For a calculation of all order corrections see \cite{Loran:2010bd})

 We start from the modular invariant   partition function of  BMSFT on a torus defined by
 \begin{equation}
 \label{s2:1}
  Z_{0}( \hat{\beta}_{u} | \hat{\beta}_{\phi} ) = \Tr e^{-\hat{\beta}_{u} (M_{0} - \frac{C_{M}}{24}) + \hat{\beta}_{\phi} (L_{0} -\frac{C_{L}}{24}) }
  = e^{\hat{\beta}_{u} \frac{C_{M}}{24} - \hat{\beta}_{\phi} \frac{C_{L}}{24} }  Z( \hat{\beta}_{u} | \hat{\beta}_{\phi} ),
 \end{equation}
where $ Z( \hat{\beta}_{u} | \hat{\beta}_{\phi} )$ is
 \begin{equation}
 \label{s2:2}
  Z( \hat{\beta}_{u} | \hat{\beta}_{\phi} ) =\Tr e^{-\hat{\beta}_{u} M_{0} + \hat{\beta}_{\phi} L_{0}}= \sum_{h_M,h_L} e^{( -\hat{\beta}_{u} h_{M} + \hat{\beta}_{\phi} h_{L})} d(h_{M},h_{L}),
 \end{equation}
 and identification of torus are
 \begin{equation}\label{simple identification} 
 (\hat u,\hat \phi)\sim(\hat u+i\hat\beta_u,\hat\phi-i\hat\beta_\phi)\sim(\hat u,\hat\phi-2\pi).
 \end{equation}
 Here, $h_L$ and $h_M$ are respectively the eigenvalues of $L_0$ and $M_0$. It is shown that the BMS modular invariant partition function satisfies \cite{Jiang:2017ecm}
\begin{equation}
\label{s2:3}
 Z_{0}( \hat{\beta}_{u} | \hat{\beta}_{\phi} ) =Z_{0}\left(-4 \pi^2 \frac{\hat{\beta}_{u}}{\hat{\beta}_{\phi}^2} | \frac{4 \pi^2}{\hat{\beta}_{\phi}}\right ). 
\end{equation}
Plugging equation \eqref{s2:3} into \eqref{s2:1} gives,
\begin{equation}
\label{s2:4}
 Z( \hat{\beta}_{u} | \hat{\beta}_{\phi} ) = e^{ \hat{\beta}_{u} \frac{C_M}{24} - \hat{\beta}_{\phi} \frac{C_L}{24} - \frac{ \pi^2 \hat{\beta}_{u} C_{M}}{6 \hat{\beta}_{\phi}^2} - \frac{ \pi^2 C_{L}}{6 \hat{\beta}_{\phi}} } Z\left(-4 \pi^2 \frac{\hat{\beta}_{u}}{\hat{\beta}_{\phi}^2} | \frac{4 \pi^2}{\hat{\beta}_{\phi}}\right ). 
\end{equation}
 By using the inverse Laplace transformation in last term of \eqref{s2:2} we find
 \begin{equation}
 \label{s2:5}
  d(h_{L} , h_{M} ) = \int d\hat{\beta}_{u} d\hat{\beta}_{\phi} e^{ \hat{\beta}_{u} \frac{C_M}{24} - \hat{\beta}_{\phi} \frac{C_L}{24} - \frac{ \pi^2 \hat{\beta}_{u} C_{M}}{6 \hat{\beta}_{\phi}^2} - \frac{ \pi^2 C_{L}}{6 \hat{\beta}_{\phi}} + \hat{\beta}_{u} h_M - \hat{\beta}_{\phi} h_L } Z\left(-4 \pi^2 \frac{\hat{\beta}_{u}}{\hat{\beta}_{\phi}^2} | \frac{4 \pi^2}{\hat{\beta}_{\phi}} \right).
 \end{equation}
In order to simplify \eqref{s2:5},  we use two approximations. First, we consider large charges which yields
 \begin{equation}
 \label{s2:6}
  d(h_{L} , h_{M} ) = \int d\hat{\beta}_{u} d\hat{\beta}_{\phi} e^{ - \frac{ \pi^2 \hat{\beta}_{u} C_{M}}{6 \hat{\beta}_{\phi}^2} - \frac{ \pi^2 C_{L}}{6 \hat{\beta}_{\phi}} + \hat{\beta}_{u} h_M - \hat{\beta}_{\phi} h_L } Z\left(-4 \pi^2 \frac{\hat{\beta}_{u}}{\hat{\beta}_{\phi}^2} | \frac{4 \pi^2}{\hat{\beta}_{\phi}}\right ).
 \end{equation}
 Then, we approximate \eqref{s2:5}   around the saddle point  given by
 \begin{align}
 \label{s2:7}
 (\hat{\beta}_{\phi}^{s})^2 =   \frac{\pi^2 C_{M}}{6 h_{M}},\qquad \hat{\beta}_{u}^{s} ={\hat{\beta}_{\phi}^{s}\over 2h_M\,C_M} (C_{M} h_{L} - C_{L} h_{M} ) .
 \end{align}
Finally, the entropy or Cardy like formula for BMSFT reads as \footnote{It is assumed that the partition function is slowly varying at the extremum.}
\begin{equation}
 \label{s2:8}
 S^{0} = \log{d(h_{L} , h_{M} )} = -{\pi^2\over 3 (\hat{\beta}_{\phi}^{s})^2 }\left(C_M \hat{\beta}_{u}^{s}+C_L \hat{\beta}_{\phi}^{s}\right).
\end{equation}
To find logarithmic correction we expand the integral \eqref{s2:6} around the saddle point \eqref{s2:7} up to quadratic term:
\begin{equation}
\label{s2:9}
   d(h_{L} , h_{M} ) = e^{S^{0}} \int d\hat{\beta}_{u} d\hat{\beta}_{\phi} e^{\frac12 (X^2 -Y^2)},
 \end{equation}
where 
\begin{align}
\label{s2:10}
 X &=  {\pi C_M\over 3 \hat{\beta}^{s}_{\phi} A}(\hat{\beta}_{u} - \hat{\beta}^{s}_{u})  , \qquad Y = {\pi A\over (\hat{\beta}^{s}_{\phi})^2} \left[ -(\hat{\beta}_{\phi} - \hat{\beta}^{s}_{\phi})+ {C_M \hat{\beta}^{s}_{\phi}\over 3 A^2} (\hat{\beta}_{u}-\hat{\beta}^{s}_{u})\right],
 \\
 \label{s2:11}
 A &= \sqrt{C_M \hat{\beta}^{s}_{u}+\frac13 C_L\hat{\beta}^{s}_{\phi}}.
\end{align}
Using \eqref{s2:10}, Eq. \eqref{s2:9} takes the form
 \begin{equation}
   d(h_{L} , h_{M} ) = e^{S^{0}}\left(-{\pi^2 C_M\over 3(\hat\beta^s_\phi)^3}\right)^{-1} \int dX dY e^{\frac{X^2 -Y^2}{2} }.
 \end{equation}
In the above equation the result of integration is just a  number. Thus we find  correction to the entropy up to a constant as \cite{Bagchi:2013qva}
\begin{equation}\label{final entropy beta}
 S = \log{ d(h_{L} , h_{M} ) } =  -{\pi^2\over 3 (\hat{\beta}_{\phi}^{s})^2 }\left(C_M \hat{\beta}_{u}^{s}+C_L \hat{\beta}_{\phi}^{s}\right)-3\log\left(-{C_M^\frac13\over \hat{\beta}_{\phi}^{s} }\right)+\text{constant}.
\end{equation}
The interesting point is that the logarithmically corrected term can be rewritten (up to a constant) as the derivative of the leading term with respect to $C_L$:
\begin{equation}
S=S_0-3\log\left(C_M^{\frac13}\dfrac{\partial S_0}{\partial C_L}\right).
\end{equation}

 \section{Logarithmic Correction to Entanglement Entropy}
 In this section, we put together the results from sections two and three to find the logarithmic correction of BMSFT entanglement entropy. As mentioned before, the idea is to map entangled states to thermal states and then calculate entropy. The entropy is computed using BMSFT Cardy-like formula \eqref{final entropy beta} which  also has a logarithmic correction. In the derivation of \eqref{final entropy beta} the identification of coordinates \eqref{simple identification} plays an essential role. It is possible to use \eqref{final entropy beta} for finding the degeneracy  of thermal states with more generic identification of coordinates as
 \begin{equation}\label{generic identification}
  (\tilde u,\tilde\phi)\sim(\tilde u+i\bar a,\tilde \phi-ia)\sim (\tilde u+2\pi \bar b,\tilde \phi-2\pi b)
  \end{equation} 
 The coordinate change between $(\tilde u,\tilde \phi)$ and $(\hat u,\hat \phi)$ is a BMS transformation,
 \begin{equation}
 \hat \phi={\tilde \phi\over b},\qquad \hat u={\tilde u \over b}+{\bar b\over b^2}\tilde \phi,
 \end{equation}
 where 
 \begin{equation}
 \hat\beta_\phi={a\over b},\qquad \hat\beta_u={\bar a\,b-a\,\bar b\over b^2}.
 \end{equation}
 Thus  Cardy-like formula \eqref{final entropy beta} can be written as
 \begin{equation}\label{entropy generic identification}
 S=-{\pi^2\over 3}\left(C_L{b\over a}+C_M{\bar a\,b-a\,\bar b\over a^2}\right)-3\log\left(-C_M^\frac{1}{3}{b\over a}\right)
 \end{equation}
In order to find the logarithmic correction of BMSFT entanglement entropy, it is enough to map the entanglement  entropy to a thermal entropy and then use \eqref{entropy generic identification}. As it was reviewed in section \ref{EE of BMSFT},  the Rindler transformation  which governs this map is determined in such a way that finally induces the  thermal identification \eqref{s1:15}. Comparing \eqref{s1:15} to \eqref{generic identification} shows that 
\begin{equation}
a=\tilde\beta_\phi,\qquad \bar a=\tilde \beta_u.
\end{equation}
 The values of $\tilde\beta_\phi$, $\tilde\beta_u$ and $b$, $\bar b$ depend on the details of the Rindler transformation. These are given in terms of the cut-offs and the interval for which entanglement entropy is calculated.  
 
Starting from a regulated interval in the BMSFT given by
\begin{equation}
 (-{l_u\over2}+\epsilon_u,-{l_\phi\over 2}+\epsilon_\phi)\to ({l_u\over2}-\epsilon_u,{l_\phi\over 2}-\epsilon_\phi),
 \end{equation} 
 the Rindler transformation yields the following results \cite{Jiang:2017ecm}:
 \begin{itemize}
 \item For the zero temperature BMSFT on the plane we have
 \begin{align}
 a&=\tilde \beta_\phi=-{2\pi^2\over \log{l_\phi\over\epsilon_\phi}},\qquad \bar a=\tilde\beta_u=-{\tilde \beta_\phi^2\over 2\pi^2}\left({l_u\over l_\phi}-{\epsilon_u\over \epsilon_\phi}\right),\\
 b&=-{\tilde\beta_\phi\over 2\pi^2}\log{l_\phi\over \epsilon_\phi},\qquad \bar b={1\over2\pi^2}\left({\tilde\beta_\phi l_u\over l_\phi}-{\tilde\beta_\phi\epsilon_u\over\epsilon_\phi}-\tilde\beta_u\log{l_\phi\over\epsilon_\phi}\right),
 \end{align}  
 Then the logarithmically corrected Cardy-like formula \eqref{entropy generic identification} becomes 
 \begin{equation}
 S_{EE}={C_L\over 6}\log{l_\phi\over\epsilon_\phi}+{C_M\over 6}\left({l_u\over l_\phi}-{\epsilon_u\over\epsilon_\phi}\right)-3\log\left(C_M^\frac{1}{3}\log{l_\phi\over\epsilon_\phi}\right)+\text{constant}.
 \end{equation}
 The third term in the above formula is the calculated correction.
 
 \item For the finite temperature BMSFT with identification
 \begin{equation}
 (u,\phi)\sim (u+i\beta_u,\phi-i\beta_\phi),
 \end{equation}
 we can use the results of \cite{Jiang:2017ecm}  and Eq. \eqref{entropy generic identification} to write
 \begin{align}
\nonumber S_{EE}={C_L\over 6}\log\left({\beta_\phi\over\pi\epsilon_\phi}\sinh{\pi l_\phi\over\beta_\phi}\right)&+{C_M\over 6}{1\over \beta_\phi}\left[\pi\left(l_u+{\beta_u\over\beta_\phi}l_\phi\right)\coth{\pi l_\phi\over \beta_\phi}-\beta_u\right]-{C_M\epsilon_u\over 6\,\epsilon_\phi}\\ &-3\log\left(C_M^{1\over3}\log\left({\beta_\phi\over \pi\epsilon_\phi}\sinh{\pi l_\phi\over \beta_\phi}\right)\right)+\text{constant}.
 \end{align}
 The forth term in the above equation is the obtained correction for finite temperature.
 
 \item For the zero temperature BMSFT on the cylinder, the entanglement  entropy with logarithmic correction reads as
 \begin{equation}
 S_{EE}={C_L\over 6}\log\left({2\over\epsilon_\phi}\sin{l_\phi\over2}\right)+{C_M\over 12}\left(l_u\cot{l_\phi\over 2}-{2\epsilon_u\over \epsilon_\phi}\right)-3\log\left(C_M^{1\over3}\log\left({2\over\epsilon_\phi}\sin{l_\phi\over2}\right)\right)+\text{costant}.
 \end{equation}
  \end{itemize} 
 It is clear from  all the above cases that up to a constant, the entanglement entropy formula including the logarithmic correction is given by
 \begin{equation}\label{final S derivative}
S_{EE}=S_0-3\log\left(C_M^{\frac13}\dfrac{\partial S_0}{\partial C_L}\right)
\end{equation}
 This formula is also valid for the thermal entropy when $S_0$ is the Cardy-like formula. Thus, we propose a universal  form for the logarithmic correction of entanglement entropy and thermal entropy. We expect the same common form for the CFT case. In the next section, we propose a similar form of logarithmic correction to CFT thermal and entanglement entropy and try to find \eqref{final S derivative} by taking limit from CFT formula.
 
 \section{Logarithmic correction of BMSFT  entropy by taking limit from CFT counterpart}
 For a two dimensional CFT with central charges $c$ and $\bar c$ and right and left temperatures $\beta$ and $\bar\beta$, the Cardy formula is 
 \begin{equation}\label{cardy formula}
  S_0={\pi^2\over 3}\left({c\over\beta}+{\bar c\over \bar \beta}\right).
  \end{equation} 
 Using AdS/CFT correspondence, this formula results in the same entropy as the outer horizon entropy of asymptotically AdS black holes in the gravity side. The logarithmic correction to \eqref{cardy formula} was evaluated in \cite{Carlip:2000nv}:
 \begin{equation}\label{cft log correction}
  S_{log}=-{3\over2}\log\left({c^{1/3}\over\beta}\right)-{3\over2}\log\left({\bar c^{1/3}\over\bar\beta}\right)+\text{constant}.
  \end{equation} 
 Employing \eqref{cardy formula} and \eqref{cft log correction} we can write
 \begin{equation}\label{CFT final entropy}
 S=S_0+S_{log}=S_0-{3\over2}\log\left(c^{1/3}\dfrac{\partial S_0}{\partial c}\right)-{3\over2}\log\left(\bar c^{1/3}\dfrac{\partial S_0}{\partial \bar c}\right)+\text{constant}.
 \end{equation}
 Using Rindler method we can expect the same formula for the logarithmically corrected CFT entanglement entropy. Thus we propose \eqref{CFT final entropy} as a universal formula that can be used for both the thermal entropy and the entanglement entropy. 
 
It is known that taking flat space limit from the asymptotically AdS spacetimes (written in the appropriate coordinate) yields asymptotically flat spacetimes. It is proposed in \cite{Bagchi:2012cy} that the flat space limit in the bulk corresponds to taking the ultra-relativistic limit in the boundary CFT. In other words, BMSFT is given by taking ultra-relativistic limit from the CFT. Starting from conformal algebra in two dimensions, one can introduce an ultra-relativistic contraction and generate BMS algebra \cite{Bagchi:2012cy}. The relation between central charges of conformal algebra and BMS algebra is given by
\begin{equation}\label{limit of central charge}
  C_L=\lim_{\epsilon\to 0}\left(c-\bar c\right),\qquad  C_M=\lim_{\epsilon\to 0}\epsilon\left(c+\bar c\right)
  \end{equation} 
where $\epsilon$ is a dimensionless parameter which corresponds to $G/\ell$ on the gravity side. $G$ is the Newton constant and $\ell$ is the AdS-radius. Thus $\epsilon\to 0$ in the boundary  corresponds to $\ell\to \infty$.

It is plausible to find all BMSFT quantities by taking limit from the CFT counterparts. Thus one can look for the possible relation between \eqref{final S derivative} and \eqref{CFT final entropy}. However, assuming $S_0$ in \eqref{CFT final entropy} as the Cardy formula \eqref{cardy formula} and using \eqref{limit of central charge} does not result in the Cardy-like formula \eqref{s2:8}. It was shown in \cite{Fareghbal:2014qga} and \cite{Riegler:2014bia} that the appropriate formula which its limit yields the Cardy-like formula is 
\begin{equation}\label{inner formula}
S_{inner}^0={\pi^2\over 3}\left({c\over\beta}-{\bar c\over \bar \beta}\right).
\end{equation}
For taking limit we should scale the temperatures as below:
\begin{equation}
\beta_u=\lim_{\epsilon\to 0}\left({\beta-\bar\beta \over 2 \epsilon}\right),  \qquad \beta_\phi=\lim_{\epsilon\to 0}-\left({\beta+\bar\beta \over 2 }\right). 
\end{equation}
 
Using AdS/CFT correspondence,  \eqref{inner formula} corresponds to the inner horizon entropy of asymptotically AdS black holes. Thus, we expect that taking limit from  \eqref{CFT final entropy} which is written for the outer horizon does not yield \eqref{final S derivative}.  In the rest of this section, we introduce appropriate formula which taking limit from it result in   \eqref{final S derivative}.

Let us start from $S^0$ in \eqref{final S derivative}. When it is  the thermal entropy given by the Cardy-like formula, the corresponding formula in the CFT will be \eqref{inner formula}. The question is then what is the corresponding CFT formula when $S^0$ in \eqref{final S derivative} is not the leading order of entanglement entropy?

It is known that taking ultra-relativistic limit from the entanglement entropy of CFT is not well-defined. For example, for a  zero temperature CFT on a plane, the entanglement entropy of an interval 
\begin{equation}
 -{R\over2}(\cosh\kappa,\sinh \kappa)\to {R\over2}(\cosh\kappa,\sinh \kappa),
 \end{equation} 
is given by \cite{Castro:2014tta}
\begin{equation}\label{outer EE}
S_{EE}={c+\bar c\over2}\log{R\over\epsilon_R}-{c-\bar c\over 6}\kappa
\end{equation}
where $\epsilon_R$ is a cut-off. Using \eqref{limit of central charge} together with ultra-relativistic contraction $t\to\epsilon t$ shows that \eqref{outer EE}  is divergent in the $\epsilon\to0$ limit.

It is proposed in \cite{Hosseini:2015uba} that BMSFT entanglement entropy can be found by taking the limit from the CFT counterpart. In the prescription of  \cite{Hosseini:2015uba}  it is argued that the symmetries are enough to fix the form of the entanglement entropy. Then they use a new contraction of conformal algebra which results in BMS algebra. But the relation between central charges of conformal algebra and BMS algebra differs from \eqref{limit of central charge}. Since the final algebra is the same as the ultra-relativistic contraction of \cite{Bagchi:2012cy}, the author of \cite{Hosseini:2015uba} argued that their results are the entanglement entropy of BMSFT.    
 
 If we continue the logic which relates the Cardy-like formula to the limit of inner horizon entropy then we conclude that the entanglement entropy of BMSFT should be related to a formula in the CFT which is transformed to the inner horizon formula by using Rindler transformation     \cite{Jiang:2017ecm}. It is not difficult to check that the  formula 
 
 \begin{equation}\label{inner EE}
S_{inner}={c-\bar c\over2}\log{R\over\epsilon_R}-{c+\bar c\over 6}\kappa
\end{equation}
 results in the leading term of entanglement entropy of zero temperature BMSFT on the plane after taking the ultra-relativistic limit. This formula is nothing but the inner horizon Cardy formula \eqref{inner formula} if we use Rindler transformation \cite{Jiang:2017ecm}.
 Therefore we conclude that $S^0$ in \eqref{final S derivative} is given by taking limit from a formula which is related to the inner horizon of dual spacetime. Similarly, the logarithmic correction in \eqref{final S derivative} is also given by taking limit from a formula which is the logarithmic correction to the inner horizon entropy. It is not difficult to check that taking $\epsilon\to 0$ limit from
 \begin{equation}\label{s log inner}
  S_{log,inner}=-{3\over2}\log\left| c^{1/3}\dfrac{\partial S_0}{\partial c}\right |-{3\over2}\log\left | \bar c^{1/3}\dfrac{\partial S_0}{\partial \bar c}\right |-\log \epsilon, 
  \end{equation} 
results in the logarithmic term of \eqref{final S derivative} up to a constant. The interpretation of \eqref{s log inner} in the gravity is of importance. As mentioned before $\epsilon$ in the field theory corresponds to $G/\ell$ on the bulk side. For the BTZ black holes, we have
\begin{equation}\label{beta of CFT}
\beta={2\pi\ell\over (r_++r_-)},\qquad \bar\beta={2\pi\ell\over (r_+-r_-)}.
\end{equation}
where $r_\pm$ are the radii of inner and outer horizons. Thus, using \eqref{inner formula}, \eqref{s log inner} and \eqref{beta of CFT} , we can find the logarithmic  correction to the inner horizon entropy of BTZ black hole as
 \begin{equation}\label{BTZ inner}
 S^{BTZ}_{inner}={\pi r_-\over 2 G}-{3\over 2}\log{r_+^2-r_-^2\over \ell^2}+\text{constant},
 \end{equation}
 where we have substituted central charges as $c=\bar c={3\ell\over 2 G}$. On the other hand, using \eqref{cardy formula}, \eqref{cft log correction} and \eqref{beta of CFT} we find that
 \begin{equation}\label{BTZ outer}
 S^{BTZ}_{outer}={\pi r_+\over 2 G}-{3\over 2}\log{r_+^2-r_-^2\over \ell^2}-\log{\ell\over G}+\text{constant}.
 \end{equation}
It is clear from \eqref{BTZ inner} and \eqref{BTZ outer} that due to the logarithmic corrections, multiplication of inner and outer horizon entropies depends not only on the angular momentum but also on the mass of BTZ. This corrects the result of \cite{Larsen:1997ge} which  in the leading term the multiplication of inner and outer entropies is mass independent.

\section{Conclusion} 

In this paper we introduced a generic formula for the logarithmic corrections of the BMSFT thermal and  the entanglement entropies. Our derivation is based on the Rindler method which makes connections between the thermal and the entanglement entropies. The most important application of the present work will reveal itself in the reconstruction of the bulk dynamics beyond the classical regime. This reconstruction has been done recently in the bulk AdS case through the perturbation of the entanglement entropy \cite{Faulkner:2013ica}, \cite{Lashkari:2013koa}. After generalizing the method of \cite{Faulkner:2013ica} for the flat-space holography, we will be able to derive the quantum corrections to the Einstein gravity  without the cosmological constant field equations using the results of the current paper. 

Most of the works in the flat-space holography can be performed by taking the flat space-limit from the AdS/CFT calculations. One of the possible roads for finding the corrections of the BMSFT entanglement entropy formula is taking limit from the calculation of \cite{Barrella:2013wja}
 which studies the one-loop bulk corrections to the Ryu-Takayanagi formula.

\subsubsection*{Acknowledgements}
The authors would like to thank S.~M.~Hosseini for his useful comments on the manuscript. We are also grateful to Yousef Izadi for his comments on the revised version. We would like to thank the referee for his/her useful comments.  This research is  supported by research grant No. 600/1476 of Shahid Beheshti University, G.C. .

\appendix



\begin{thebibliography}{}

\bibitem{Maldacena:1997re} 
  J.~M.~Maldacena,
  ``The Large N limit of superconformal field theories and supergravity,''
  Int.\ J.\ Theor.\ Phys.\  {\bf 38}, 1113 (1999)
  [Adv.\ Theor.\ Math.\ Phys.\  {\bf 2}, 231 (1998)]
  doi:10.1023/A:1026654312961
  [hep-th/9711200].


\bibitem{BMS}
H.~Bondi, M.~G. van~der Burg, and A.~W. Metzner, ``Gravitational
waves in general relativity. 7. {W}aves from axisymmetric isolated systems,'' {\em
  Proc.\ Roy.\ Soc.\ Lond. A} {\bf 269} (1962)
21.

R.~K. Sachs, ``Gravitational waves in general relativity. 8.
{W}aves in asymptotically flat space-times,'' {\em Proc.\ Roy.\ Soc.\ Lond.\ A} {\bf
  270} (1962)
103.

R.~K. Sachs, ``Asymptotic symmetries in gravitational theory,''
{\em Phys.
  Rev.} {\bf 128} (1962) 2851.

 \bibitem{Ashtekar:1996cd} 
  A.~Ashtekar, J.~Bicak and B.~G.~Schmidt,
  ``Asymptotic structure of symmetry reduced general relativity,''
  Phys.\ Rev.\ D {\bf 55}, 669 (1997)
  [gr-qc/9608042].



\bibitem{Barnich:2006av} 
  G.~Barnich and G.~Compere,
  ``Classical central extension for asymptotic symmetries at null infinity in three spacetime dimensions,''
  Class.\ Quant.\ Grav.\  {\bf 24}, F15 (2007)
  [gr-qc/0610130].

 G.~Barnich and C.~Troessaert,
 ``Symmetries of asymptotically flat 4 dimensional spacetimes at null infinity revisited,''
  arXiv:0909.2617 [gr-qc].

G.~Barnich and C.~Troessaert,
  ``Aspects of the BMS/CFT correspondence,''
  JHEP {\bf 1005}, 062 (2010)
  [arXiv:1001.1541 [hep-th]].






\bibitem{Bagchi:2010zz}
  A.~Bagchi,
  ``Correspondence between Asymptotically Flat Spacetimes and Nonrelativistic Conformal Field Theories,''
  Phys.\ Rev.\ Lett.\  {\bf 105}, 171601 (2010).

  A.~Bagchi,
  ``The BMS/GCA correspondence,''
  arXiv:1006.3354 [hep-th].
%
%
%
\bibitem{Bagchi:2012cy} 
  A.~Bagchi and R.~Fareghbal,
  ``BMS/GCA Redux: Towards Flatspace Holography from Non-Relativistic Symmetries,''
  [arXiv:1203.5795 [hep-th]].
  
  



\bibitem{Bagchi:2014iea} 
  A.~Bagchi, R.~Basu, D.~Grumiller and M.~Riegler,
  ``Entanglement entropy in Galilean conformal field theories and flat holography,''
  Phys.\ Rev.\ Lett.\  {\bf 114}, no. 11, 111602 (2015)
  doi:10.1103/PhysRevLett.114.111602
  [arXiv:1410.4089 [hep-th]].



%
\bibitem{Calabrese:2004eu} 
  P.~Calabrese and J.~L.~Cardy,
  ``Entanglement entropy and quantum field theory,''
  J.\ Stat.\ Mech.\  {\bf 0406}, P06002 (2004)
  doi:10.1088/1742-5468/2004/06/P06002
  [hep-th/0405152].
%



\bibitem{Ryu:2006bv} 
  S.~Ryu and T.~Takayanagi,
  ``Holographic derivation of entanglement entropy from AdS/CFT,''
  Phys.\ Rev.\ Lett.\  {\bf 96}, 181602 (2006)
  doi:10.1103/PhysRevLett.96.181602
  [hep-th/0603001].




\bibitem{Witten:1988hc} 
  E.~Witten,
  ``(2+1)-Dimensional Gravity as an Exactly Soluble System,''
  Nucl.\ Phys.\ B {\bf 311}, 46 (1988).
  doi:10.1016/0550-3213(88)90143-5


\bibitem{Hosseini:2015uba} 
  S.~M.~Hosseini and Á.~Véliz-Osorio,
  ``Gravitational anomalies, entanglement entropy, and flat-space holography,''
  Phys.\ Rev.\ D {\bf 93}, no. 4, 046005 (2016)
  doi:10.1103/PhysRevD.93.046005
  [arXiv:1507.06625 [hep-th]].




\bibitem{Basu:2015evh} 
  R.~Basu and M.~Riegler,
  ``Wilson Lines and Holographic Entanglement Entropy in Galilean Conformal Field Theories,''
  Phys.\ Rev.\ D {\bf 93}, no. 4, 045003 (2016)
  doi:10.1103/PhysRevD.93.045003
  [arXiv:1511.08662 [hep-th]].




  
 
  
\bibitem{Jiang:2017ecm} 
  H.~Jiang, W.~Song and Q.~Wen,
  ``Entanglement Entropy in Flat Holography,''
  arXiv:1706.07552 [hep-th].


\bibitem{Casini:2011kv} 
  H.~Casini, M.~Huerta and R.~C.~Myers,
  ``Towards a derivation of holographic entanglement entropy,''
  JHEP {\bf 1105}, 036 (2011)
  doi:10.1007/JHEP05(2011)036
  [arXiv:1102.0440 [hep-th]].


\bibitem{Bagchi:2012xr} 
  A.~Bagchi, S.~Detournay, R.~Fareghbal and J.~Simón,
  ``Holography of 3D Flat Cosmological Horizons,''
  Phys.\ Rev.\ Lett.\  {\bf 110}, no. 14, 141302 (2013)
  doi:10.1103/PhysRevLett.110.141302
  [arXiv:1208.4372 [hep-th]].




 
\bibitem{Bagchi:2013qva} 
  A.~Bagchi and R.~Basu,
  ``3D Flat Holography: Entropy and Logarithmic Corrections,''
  JHEP {\bf 1403}, 020 (2014)
  doi:10.1007/JHEP03(2014)020
  [arXiv:1312.5748 [hep-th]].







\bibitem{Carlip:2000nv} 
  S.~Carlip,
  ``Logarithmic corrections to black hole entropy from the Cardy formula,''
  Class.\ Quant.\ Grav.\  {\bf 17}, 4175 (2000)
  doi:10.1088/0264-9381/17/20/302
  [gr-qc/0005017].
  
\bibitem{Sen:2012cj} 
  A.~Sen,
  ``Logarithmic Corrections to Rotating Extremal Black Hole Entropy in Four and Five Dimensions,''
  Gen.\ Rel.\ Grav.\  {\bf 44}, 1947 (2012)
  doi:10.1007/s10714-012-1373-0
  [arXiv:1109.3706 [hep-th]].  
  
  
  
\bibitem{Sen:2012dw} 
  A.~Sen,
  ``Logarithmic Corrections to Schwarzschild and Other Non-extremal Black Hole Entropy in Different Dimensions,''
  JHEP {\bf 1304}, 156 (2013)
  doi:10.1007/JHEP04(2013)156
  [arXiv:1205.0971 [hep-th]].
  
  
  
  
  
  
   
  
  
  
  
%

\bibitem{Faulkner:2013ica} 
  T.~Faulkner, M.~Guica, T.~Hartman, R.~C.~Myers and M.~Van Raamsdonk,
  ``Gravitation from Entanglement in Holographic CFTs,''
  JHEP {\bf 1403}, 051 (2014)
  doi:10.1007/JHEP03(2014)051
  [arXiv:1312.7856 [hep-th]].
  
\bibitem{Lashkari:2013koa} 
  N.~Lashkari, M.~B.~McDermott and M.~Van Raamsdonk,
  ``Gravitational dynamics from entanglement 'thermodynamics',''
  JHEP {\bf 1404}, 195 (2014)
  doi:10.1007/JHEP04(2014)195
  [arXiv:1308.3716 [hep-th]].
  
  
\bibitem{Allahbakhshi:2013rda} 
  D.~Allahbakhshi, M.~Alishahiha and A.~Naseh,
  ``Entanglement Thermodynamics,''
  JHEP {\bf 1308}, 102 (2013)
  doi:10.1007/JHEP08(2013)102
  [arXiv:1305.2728 [hep-th]].












%

\bibitem{Fareghbal:2014qga} 
  R.~Fareghbal and A.~Naseh,
  ``Aspects of Flat/CCFT Correspondence,''
  Class.\ Quant.\ Grav.\  {\bf 32}, 135013 (2015)
  doi:10.1088/0264-9381/32/13/135013
  [arXiv:1408.6932 [hep-th]].



\bibitem{Riegler:2014bia} 
  M.~Riegler,
  ``Flat space limit of higher-spin Cardy formula,''
  Phys.\ Rev.\ D {\bf 91}, no. 2, 024044 (2015)
  doi:10.1103/PhysRevD.91.024044
  [arXiv:1408.6931 [hep-th]].

 








\bibitem{Larsen:1997ge} 
  F.~Larsen,
  ``A String model of black hole microstates,''
  Phys.\ Rev.\ D {\bf 56}, 1005 (1997)
  doi:10.1103/PhysRevD.56.1005
  [hep-th/9702153].

\bibitem{Detournay:2012ug} 
  S.~Detournay,
  ``Inner Mechanics of 3d Black Holes,''
  Phys.\ Rev.\ Lett.\  {\bf 109}, 031101 (2012)
  doi:10.1103/PhysRevLett.109.031101
  [arXiv:1204.6088 [hep-th]].

\bibitem{Barnich:2012xq} 
  G.~Barnich,
  ``Entropy of three-dimensional asymptotically flat cosmological solutions,''
  JHEP {\bf 1210}, 095 (2012)
  doi:10.1007/JHEP10(2012)095
  [arXiv:1208.4371 [hep-th]].



\bibitem{Basu:2017aqn} 
  R.~Basu, S.~Detournay and M.~Riegler,
  ``Spectral Flow in 3D Flat Spacetimes,''
  arXiv:1706.07438 [hep-th].
  
  

\bibitem{Loran:2010bd} 
  F.~Loran, M.~M.~Sheikh-Jabbari and M.~Vincon,
  ``Beyond Logarithmic Corrections to Cardy Formula,''
  JHEP {\bf 1101}, 110 (2011)
  doi:10.1007/JHEP01(2011)110
  [arXiv:1010.3561 [hep-th]].



\bibitem{Castro:2014tta} 
  A.~Castro, S.~Detournay, N.~Iqbal and E.~Perlmutter,
  ``Holographic entanglement entropy and gravitational anomalies,''
  JHEP {\bf 1407}, 114 (2014)
  doi:10.1007/JHEP07(2014)114
  [arXiv:1405.2792 [hep-th]].

\bibitem{Barrella:2013wja} 
  T.~Barrella, X.~Dong, S.~A.~Hartnoll and V.~L.~Martin,
  ``Holographic entanglement beyond classical gravity,''
  JHEP {\bf 1309}, 109 (2013)
  doi:10.1007/JHEP09(2013)109
  [arXiv:1306.4682 [hep-th]].




%
%
%
%
%

%
%

%


  
\end{thebibliography}
\end{document}